\documentstyle[sprocl_my,epsf]{article}

\def\Journal#1#2#3#4{{#1} {\bf #2}, #3 (#4)}

\def\NPB{{\em Nucl. Phys.} B}
\def\PLB{{\em Phys. Lett.}  B}

\def\PRD{{\em Phys. Rev.} D}

\def\APP{{\em Acta Phys. Polon.}}
\def\SNP{{\em Sov. J. Nucl. Phys.}}

\def\be{\begin{equation}}
\def\ee{\end{equation}}
\def\bea{\begin{eqnarray}}
\def\eea{\end{eqnarray}}

\def\lsim{\mathrel{\raise.3ex\hbox{$<$\kern-.75em\lower1ex\hbox{$\sim$}}}}
\def\gsim{\mathrel{\raise.3ex\hbox{$>$\kern-.75em\lower1ex\hbox{$\sim$}}}}

\begin{document}


\title{TOP PHYSICS AT THRESHOLD AND BEYOND 
}

\author{ A. H. Hoang}

\address{
   Theory Division, CERN, CH-1211 Geneva 23, Switzerland }



\thispagestyle{empty}
\begin{minipage}{5.0in}
\begin{center}

\begin{flushright}
{\bf CERN-TH/99-271}\\
{\bf hep-ph/9909414}\\
{\bf September 1999}\\
\end{flushright}
\vspace{1.0cm}
\begin{center}
  \begin{Large}\bf
Top Physics at Threshold and Beyond
  \end{Large}
  \vspace{1.5cm}

\begin{large}
 A.~H.~Hoang
\end{large}

\vspace{.5cm}
\begin{it}
Theory Division, CERN,\\
   CH-1211 Geneva 23, Switzerland\\[.5cm]
\end{it}

  \vspace{2.5cm}
  {\bf Abstract}\\
\vspace{0.3cm}

\noindent
\begin{minipage}{4.7in}
\begin{small}
A review on theoretical aspects of top quark physics at the Linear
Collider is given with focus on the process $e^+e^-\to t\bar t$ and the
presentations given at this conference. 
\end{small}
\end{minipage}
\end{center}
\setcounter{footnote}{0}
\renewcommand{\thefootnote}{\arabic{footnote}}
\vspace{2.5cm}
\begin{it}
Plenary talk given at World-Wide Study of \\
Physics and Detectors for 
Future Linear Colliders (LCWS 99), \\
Sitges, Barcelona, Spain, 
28 Apr - 5 May 1999.
\end{it}
\vspace{3cm}
\begin{flushleft}
{\bf CERN-TH/99-271}\\
{\bf September 1999}
\end{flushleft}
\end{center}
\end{minipage}

\newpage
\pagestyle{plain}
\setcounter{page}{1}


\maketitle\abstracts{
A review on theoretical aspects of top quark physics at the Linear
Collider is given with focus on the process $e^+e^-\to t\bar t$ and the
presentations given at this conference. 
}
  
\section{Introduction}

Top quark physics constitutes one of the main tasks at the Linear
Collider (LC). Top physics at the LC is in many respects complementary
to top physics at the LHC. Although the LC has smaller statistics, it
provides a much cleaner environment, which leads to smaller systematic
uncertainties. One obvious reason is the fact that the top pair is
produced in a colour singlet state rather than in an octet like at
the LHC. 
Thus top physics at the LC can be expected to lead to
results at a high level of precision. The basic property which makes
precision studies of the top quark {\it per se} at all possible is its
large mass: in the Standard Model (SM) the top decay width is
dominated by the decay into a $b$ quark and a $W$ boson and reads
$\Gamma_t=(G_F/\sqrt{2})(M_t^3/8 \pi)\approx
1.5\,\mbox{GeV}$ at the Born level and in the limit of vanishing $W$
mass. Because the width is much larger than the typical hadronisation
scale, the top quark decays before hadronisation effects set in. This
fact makes top meson spectroscopy impossible, but, at the same time,
leads to a suppression of nonperturbative effects in top production
and decay in any kinematic regime~\cite{Bigi1,Khoze1}. In many (but
not all) respects, the 
top quark can be considered as a real particle, and properties
like its polarisation are measurable observables, which can be
determined from distributions of the top decay products. 
In addition, this feature allows the theorists to use perturbative methods to
describe the top quark a high degree of precision~\cite{Bigi1,Khoze1}. 

In this talk I review some theoretical aspects of top quark physics at
the LC focusing on the process $e^+e^-\to t\bar t$ and the presentations
given at this
conference. The presentation is subjective and not all issues of
interest can be mentioned. 
However, I hope that this talk can reflect some of the flavour of
the rich and interesting top phenomenology at the LC. 

\section{Threshold and Continuum}

Because the c.m. energy of the $e^+e^-$ collision is well known at the
GeV level it is possible to resolve the $t\bar t$ threshold
regime where Coulomb-like binding affects the $t\bar t$
dynamics. This allows to do top physics in two completely different
theoretical settings. In the continuum far above the $t\bar t$
threshold ($\sqrt{s}\gsim 2M_t+15\,\mbox{GeV}$) conventional
perturbative methods can be employed, whereas close to threshold
($\sqrt{s}\approx 2M_t$) resummations of terms $\propto\alpha_s/v$, $v$
being the c.m. top velocity, have to be carried out to all orders in
$\alpha_s$. Practically all top quark properties can be measured in
the continuum as well as in the threshold regime. The different
interplay of the top quark properties with QCD in the two regimes
allows for complementary tests of the SM and, in particular, 
of the strong interaction.

\section{Top Mass}

The top mass affects the relation between the electroweak precision
observables indirectly through loop effects. In the parameter  
$\Delta r$, which relates $M_W, M_Z, \alpha_{em}$ and $G_F$,
the top mass enters quadratically. The expected reductions of the
uncertainties of quantities like the W mass and the weak mixing angle
($\delta M_W=15(8)\,\mbox{MeV}$, $\delta \sin^2\theta_W=18(1)\times
10^{-5}$ at LHC/LC (Giga Z))~\cite{Wilson1} make it desirable to
determine the top 
mass as accurate as possible in order to improve the sensitivity to
the Higgs mass or non-SM loop effects which enter $\Delta r$ less
strongly than the top mass~\cite{Heinemeyer1}. Stringent bounds on the
Higgs mass will provide a test of the SM Higgs mechanism.
At the LC the top mass can be
measured at the per mille level in two ways. The standard method is to
reconstruct the top 
invariant mass distribution. Because systematic experimental effects
(i.e. jet energy resolution, beam effects, gluon radiation) are 
quite well understood in the $e^+e^-$ environment it will be possible
to determine the peak of the distribution to a few hundred
MeV~\cite{Bagliesi1} 
(compared to about 2 GeV at the LHC). Studies using dilepton
events~\cite{Yeh1} have shown that the peak can be determined to 200
MeV. However, one has to keep in mind that, at present, it is not
known how to relate the peak of the reconstructed invariant mass
distribution to a theoretically clean quark mass
definition. Intuitively the peak is most closely related to the pole
mass, but the latter has an intrinsic theoretical ambiguity of order
$\Lambda_{QCD}\approx$~200-300 MeV, also known as the ``pole mass
renormalon'' problem. Related but not equivalent to this problem are
QCD interconnection effects which arise from the colour
rearrangement among the top and antitop decay products in the
hadroformation process. The modelling of this phenomenon could lead to
uncertainties in the peak of around 100 MeV~\cite{Sjostrand1}. In summary
one can say that the peak in the invariant mass
distribution is ambiguous to an amount of order $\Lambda_{QCD}$
because a) the invariant mass (or the momentum) of a coloured particle is
ambiguous, i.e. its exact meaning depends on the reconstruction
method, and b) because some aspects of the colour rearrangement in the
hadroformation process are not understood yet.  
More theoretical studies of this interesting subject have to be
carried out to fully exploit the potential of the mass reconstruction
method at the LC. 

\begin{figure}[t!] 
\begin{center}
\leavevmode
\epsfxsize=2.8cm
\leavevmode
\epsffile[220 580 420 710]{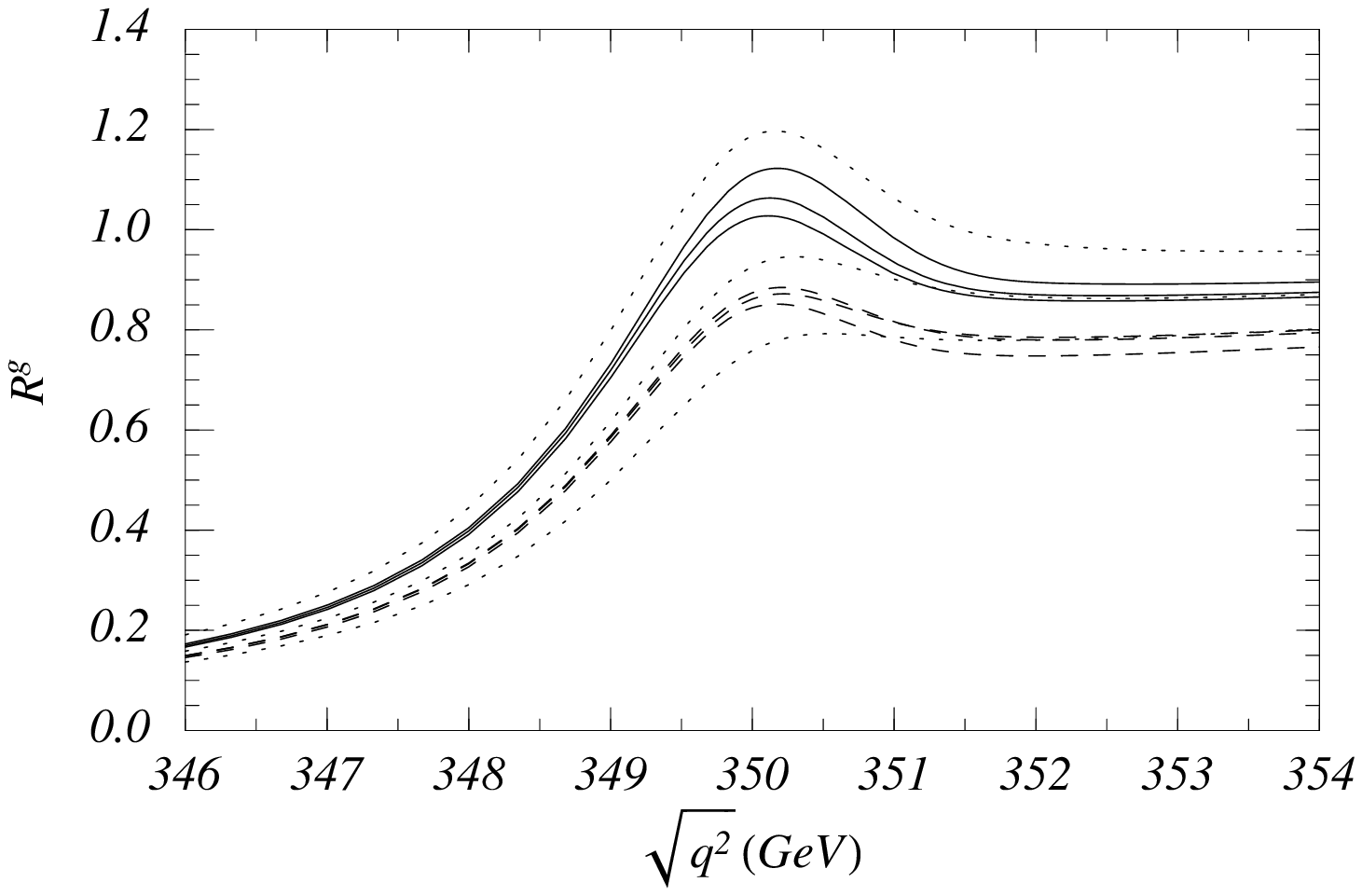}
\hspace{3cm}
\epsfxsize=2.8cm
\leavevmode
\epsffile[220 580 420 710]{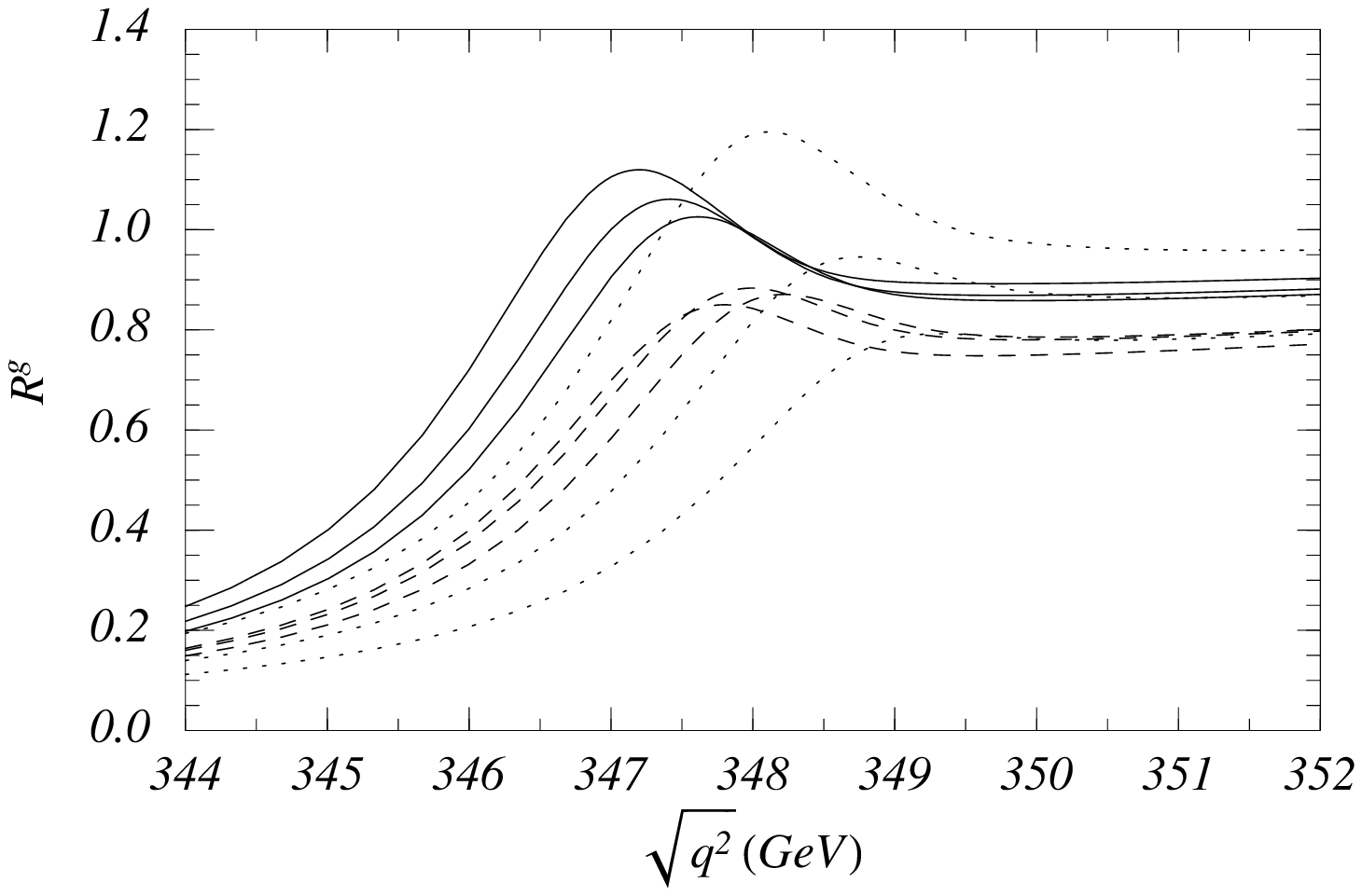}
 \end{center}
%
%
\vskip  1.7cm
 \caption{\label{figttbar}
The total photon-induced $t\bar t$ cross section devided by the point
cross section at the LC versus the c.m. energy in the threshold regime
at LO (dotted curves), NLO (dashed) and NNLO (solid) in the $1S$
(left figure) and the pole (right figure) mass schemes for
$\alpha_s(M_Z)=0.118$ and $\mu=15$, $30$, $60$~GeV. The plots have 
been taken from Ref.${}^{11}$.\mbox{\hspace{6cm}} 
}
\end{figure}

The second possibility to determine the top mass comes from a scan of
the total $t\bar t$ cross section line-shape around the $t\bar t$
threshold. The rise and the shape of the cross section can be directly
related to the top quark mass. The advantage of the threshold scan
is that the total cross section describes the production rate of
colour-singlet $t\bar t$ pairs. Thus the conceptual limitations and
problems related to the top as a coloured particle and to the colour
flow among the top decay products only play a minor (but not
negligible) role. Just recently next-to-next-to-leading order 
(NNLO) QCD calculations for the total cross section have been carried out
using the concept of effective field
theories~\cite{Teubner1,ttbarthresh,nnlottbar,Beneke1,Hoang2}. In this
approach the hierarchy $M_t\gg M_t v\,\mbox{($t$ momentum)}\,\gg M_t
v^2\,\mbox{($t$ kinetic energy)}\, > \Gamma_t\gg \Lambda_{QCD}$ 
is used to integrate out the dynamical degrees of freedom associated
with the scales $M_t$ and $M_t v$, and to derive field equations
describing the $t\bar t$ dynamics. The NNLO calculations 
demonstrate the inadequacy of the pole mass definition as it leads to
NNLO corrections in the peak position which are as large at the NLO
ones (see the right picture in Fig.~\ref{figttbar} and
also Ref.~\cite{Hoang1}). Several alternative mass definitions have been
proposed which (on the conceptual side) avoid the
$\Lambda_{QCD}$-ambiguity of the pole mass and (on the practical side)
lead to a reduced correlation of the threshold line-shape to the choice
of theoretical parameters like the renormalisation scale or 
$\alpha_s$. In the left picture in Fig.~\ref{figttbar} the total cross
section is displayed using the so called $1S$ top mass
definition~\cite{Hoang2,Teubner1}.  
(Similar results have been obtained with the so called $PS$
mass~\cite{Beneke1}.) The advantage of these alternative mass definitions is
twofold: they allow for smaller theoretical uncertainties in the
mass determination, and they can be related to the
$\overline{\mbox{MS}}$ mass, the preferred mass definition in the high
energy continuum, more reliably than the pole mass. Simulation
studies~\cite{Peralta1} 
have shown that the $1S$ and $PS$ masses can be determined with
theoretical and experimental uncertainties of around 100 MeV despite the 
beamstrahlung effects, which lead to an additional smearing of the
cross section, and the remaining theoretical uncertainties in the
normalisation of the total cross section.

A measurement of the $\overline{\mbox{MS}}$ top mass at the level of 5
GeV can be achieved if the ratio of $t\bar t g$ events versus all
$t\bar t$ events could be measured to $1\%$ at at c.m. energy of 1
TeV~\cite{Brandenburg1}. This
is possible because the $t\bar t g$ cross section also depends on the
jet-resolution scale.
For small jet-resolution scales the dependence on the top mass is enhanced.  
Calculations of the $t\bar t g$ versus $t\bar t$ ratio have been carried
out at order $\alpha_s^2$ and have shown that the order $\alpha_s^2$
corrections are of order 30\%~\cite{Brandenburg1}. Thus the calculation
of higher order calculations is required. A determination of the top
mass from the $t\bar t g$ versus $t\bar t$ ratio would not be able to
compete in precision with the mass determination from the invariant
mass distribution or the threshold scan, but it would serve as a
cross check for the previously mentioned methods.

\section{Strong Coupling}

The strong coupling governs the Coulombic attraction of the $t\bar t$
pair in the threshold regime. The resummations of terms
$\propto\alpha_s/v$ to all orders lead to a strong dependence of the
normalisation of the threshold cross section on 
$\alpha_s$, $\sigma_{t\bar t}\sim|\Psi_{t\bar t}(0)|^2\sim
\alpha_s^3$. Thus one might conclude that the threshold scan would be
a reliable way to determine the strong coupling.
The newly available NNLO corrections for the threshold cross
section, however, also reveal that the uncertainty in the
normalisation of the cross section is still significant, at least at
the order 10\%~\cite{Teubner1,ttbarthresh,nnlottbar,Beneke1,Hoang2} 
(see Fig~\ref{figttbar}). 
Taking the size of the NNLO normalisation corrections as an 
uncertainty leads to
an uncertainty in $\alpha_s(M_Z)$ of $0.012$, which is five times
larger than the combined systematic and statistical experimental
error~\cite{Peralta1}. (The present normalisation uncertainty of the
cross section at threshold also makes a measurement of the Higgs mass
or the top  Yukawa coupling from the cross section impossible.
The Higgs mass affects the cross section through electroweak 
corrections to the $t\bar t$ production vertex and through a Yukawa  
interaction potential. For $M_h$ around 100~GeV the effects of
the Yukawa potential are negligible and the Higgs effects in the
vertex corrections are at the level of several percent and quickly
decrease if $M_h$ is larger~\cite{Harlander1}.) 
This shows that a better understanding of the
normalisation of the threshold cross section is mandatory, which can
probably only be achieved by a calculation of all N${}^3$LO
corrections. In principle, $\alpha_s$ can also be determined from more
differential threshold quantities like the top three-momentum
distribution or the angular distribution. The NNLO corrections for those
quantities are not completed yet, and are expected to be
sizeable. Of particular interest in
this respect is 
the problem of nonfactorizable corrections, which come from the
exchange of gluons among the top and the top decay products. 
This problem is closely related to the colour reconnection problem
mentioned before. 

A study has also been carried out on the $\alpha_s$-determination from
the ratio $\sigma_{t\bar t}/\sigma_{\mu^+\mu^-}$ above 
threshold~\cite{Bernreuther1}. For c.m. energies above
0.5 TeV theoretical uncertainties are below 0.5\%. The uncertainty in
a determination of $\alpha_s(M_Z)$ is then dominated by the luminosity 
measurement. Assuming an uncertainty of about 2\% for the luminosity
leads to an error in $\alpha_s(M_Z)$ of about 0.007. This could serve
as a cross check for other $\alpha_s$-determinations~\cite{Burrows1}.

The ratio of $t\bar t g$ events versus all $t\bar t$ events might also
be used as a means to determine $\alpha_s$. The uncertainties of
the order $\alpha_s^2$ calculations known at present~\cite{Brandenburg1} do,
however, not allow to draw any definite conclusions.

\section{Top Yukawa Coupling}

By the time when the LC starts operation the LHC will probably have
already discovered the Higgs boson and determined its mass. A direct
measurement of the top Yukawa coupling $g_{tth}$ will then provide an
important test whether the SM Higgs mechanism, which leads to
$g_{tth}^2=\sqrt{2}G_F M_t^2$, is indeed realized for the quark mass
generation. Deviations from the SM value might indicate a different
mass generation mechanism, or could be a reflection of an extended
Higgs sector, which would make the Yukawa couplings also depend on the
Higgs mixing angles. For a light Higgs ($M_h< 2 M_t$), the situation
which is favoured by supersymmetric models and which has been mainly
studies up to now, the reaction $e^+e^-\to t\bar t h$ 
is best suited for a direct measurement of $g_{tth}$. The experimental
signature $WWbbbb$ with 6 or 8 jets is spectacular. Complete order
$\alpha_s$ calculations exist in the SM and the minimal supersymmetric
SM (MSSM)~\cite{Reina1}. The order $\alpha_s^2$ corrections are
expected to 
be small. In general, the cross section is below a few $fb$,
which makes collider designs with high luminosity
desirable. Simulations have shown~\cite{Juste1} that for the 6 and the
8 jet mode relative (combined statistical and systematical)
uncertainties in $g_{tth}$ below 10\% can be achieved for a Higgs mass
around 120~GeV at $\sqrt{s}=$800 GeV and an integrated luminosity of
1000~$fb^{-1}$. Corresponding LHC studies yield larger
uncertainties. For a heavy Higgs ($M_h>2M_t$) the modes $e^+e^-\to
Zh(\to Z t\bar t)$ and $e^+e^-\to \nu\bar\nu h(\to \nu\bar\nu t\bar
t)$ are dominant. 

\section{Anomalous Couplings, CP violation, Correlations}

The clean LC environment allows for many different ways to test the
charged or neutral current top quark couplings for non-SM or
CP-violating contributions. The possibility to change the
electron-positron beam polarisations can enhance the sensitivity of 
observables to those effects. Studies have been carried out for
CP-even observables like lepton energy spectra~\cite{Hioki1} 
or the gluon energy spectrum in $e^+e^-\to t\bar t g$~\cite{Rizzo1}, 
CP-odd asymmetries and CP-odd spin-momentum
correlations of the top decay products. Measurements are possible at
threshold and in the continuum, although statistics at threshold 
will be somewhat worse because only a relative small amount of
luminosity will be spent there. In general, sensitivities at the level
of a several to ten percent could be achieved, but a high luminosity is
needed to reach interesting sensitivities for many models.

Although CP-violation is implemented into the SM through phases in the
CKM matrix elements, it is practically impossible to detect SM
CP-violation in top physics. This is a consequence of the GIM
mechanism, which is particularly effective owing to the large top quark
mass. (For exactly the same reason is B physics very much suitable to
measure the CKM phase.)
Observed CP-violating effects in observables related to the top quark
would be a clear signal of
new physics. On the other hand, the CKM phase is unlikely to be the
only source of CP-violation in baryogenesis. Of particular interest is
CP-violation which originates from the Higgs sector~\cite{Soni1}. It would
have good chances to be detected in top physics because the top Yukawa
couplings are enhanced by the large top quark mass. In
multi-Higgs-doublet models (MHDM's) CP-violation can be either
implemented explicitly or, in models with more than two Higgs
doublets, arises spontaneously. In MHDM's the top Yukawa couplings
could also be further enhanced due to their dependence on the Higgs mixing
angles. In 3HDM's CP-odd $\tau$ transverse polarisation asymmetries
in the Higgs decay $t\to b\tau\nu$ could reach order 10\%.
An observation would signal CP-violation in the charged Higgs
sector. In certain 2HDM's CP-violating phases in the neutral sector
could be  
detected for small values of $\tan\beta$(=ratio of the two VEV's) in
the processes $e^+e^-\to t\bar t h$ and $e^+e^-\to t\bar t Z$ using
momentum correlations and optimised observables~\cite{Soni1}. For
$M_h>2M_t$ and known Higgs mass it might also be possible to reconstruct
the decay $h\to t\bar t$ from 
$e^+e^-\to t\bar t Z$~\cite{Bernreuther2} or from  
$e^+e^-\to t\bar t \nu_e\bar\nu_e$~\cite{Atwood1}. For sufficiently high
luminosity CP-violating phases could then be measured in spin-momentum
correlations.

\section{Top Decay}

Apart from the standard final state reconstruction method,
a promising way for a direct determination of the top quark width
could be the threshold scan or the measurement of the forward-backward
asymmetry in top quark production close to threshold. The peak of the
total cross section is more pronounced for smaller top width, whereas
the width dependence of the forward-backward asymmetry originates from 
the overlap of $t\bar t$ S-wave and P-wave amplitudes which 
is bigger for larger top width. At present, however, no
definite conclusions can be drawn in view of the (potentially) large NNLO
corrections. In addition, the problems of unambiguously defining the
direction of flight of the top as a coloured particle and of properly
understanding the effects of nonfactorizable
corrections need to be addressed.
Putting the problems of uncalculated higher order corrections and the
conceptual issues just mentioned aside, experimental uncertainties of
10-20\% could be achieved from the threshold~\cite{Fujii1}. 
A method to extract the top width from the interference of decay and
production stage radiation of gluons in the process $e^+e^-\to t\bar t
g$ has been investigated~\cite{Orr1}. However, no definite conclusions
can yet be drawn here either. 

Due to the strong CKM and GIM suppressions any observed top decay
other than into a bottom quark and a W boson would practically imply
non-SM physics. In is therefore interesting to examine non-SM decay
modes of the top quark. For the LC we have to keep in mind that
branching ratios for rare decays must be larger than order $10^{-5}$
to be visible. The work on this rich subject has been extensive for
the LC and also for the LHC and only a few impressions can be given
here. In supersymmetric extensions of the SM top decays into charged
Higgs bosons ($t\to b H^+$) and into stops and neutralinos
($t\to\tilde t\tilde\chi^0_1$) or sbottoms and charginos
($t\to\tilde b\tilde\chi^+_1$) are possible. If kinematically allowed,
those modes can have branching ratios of more than 10\% using the
present constraints on supersymmetric parameters. The observation of
$t\to b H^+$ would not necessarily be a hint for supersymmetry because
this mode also exists in general multi-Higgs
models. In the framework of the MSSM the FCNC decay $t\to X c$ is
dominated by final states containing neutral Higgs bosons and could
reach a branching ratio~\cite{Sola1} of several $10^{-4}$.

\section{Summary}
The prospects for the top mass determination at the level of 100-200
MeV 
at the LC are very good. The mass determination from the threshold
scan of the total $t\bar t$ cross section line-shape seems well
understood. Some more conceptual progress has to be 
achieved to control the mass reconstruction method at the same level. 
At present it seems difficult to achieve
measurements of $\alpha_s$ from top observables which could compete
with other methods, and probably the best way to proceed is to take
$\alpha_s$ as an input from somewhere else. A better understanding of
the normalisation of the $t\bar t$ threshold cross section is needed
improve this situation. The determination of the top Yukawa coupling is
well studied for the case of a light Higgs for the reference process
$e^+e^-\to t\bar t h$; uncertainties below 10\%
seem realistic from 6 and 8 jet modes. CP-violating phases from models
with extended Higgs sectors can be accessible for sufficient luminosity. 
The physics of non-standard top decays is very rich. A number of rare
decay modes that could be visible at the LC exist in extensions of the
SM.

\section*{Acknowledgements}

This work is supported in part by the EU Fourth Framework Program
``Training and Mobility of Researchers'', Network ``Quantum
Chromodynamics and Deep Structure of Elementary Particles'', contract
FMRX-CT98-0194 (DG12-MIHT).

\end{document}